# Structural diversity and the role of particle shape and dense fluid behavior in assemblies of hard polyhedra

Pablo F. Damasceno[1*], Michael Engel[2*], Sharon C. Glotzer[1,2,3†]

[1]Applied Physics Program, [2]Department of Chemical Engineering, and [3]Department of Materials Science and Engineering, University of Michigan, Ann Arbor, Michigan 48109, USA.

* These authors contributed equally.

† Corresponding author: sglotzer@umich.edu

**A fundamental characteristic of matter is its ability to form ordered structures under the right thermodynamic conditions. Predicting these structures – and their properties – from the attributes of a material's building blocks is the holy grail of materials science. Here, we investigate the self-assembly of 145 hard convex polyhedra whose thermodynamic behavior arises solely from their anisotropic shape. Our results extend previous works on entropy-driven crystallization by demonstrating a remarkably high propensity for self-assembly and an unprecedented structural diversity, including some of the most complex crystalline phases yet observed in a non-atomic system. In addition to 22 Bravais and non-Bravais crystals, we report 66 plastic crystals (both Bravais and topologically close-packed), 21 liquid crystals (nematic, smectic, and columnar), and 44 glasses. We show that from simple measures of particle shape and local order in the disordered fluid, the class of ordered structure can be predicted.**

**Introduction**

Conventional matter exhibits a rich diversity of structure, from simple to complex crystals and quasicrystals in atomic systems, to complex crystals, liquid crystals, and plastic crystals in molecular materials. Despite great progress in the fabrication of nanoparticle and colloidal crystals through the use of binary mixtures of oppositely charged particles[1-3], DNA linkers[4-6], anisotropic shape[7-11], and depletion interactions[12-14], the structural diversity exhibited by these assemblies lags far behind that of traditional materials.

For hard colloids where the only interaction is that of the excluded volume between particles, the diversity of structure is even lower, but growing. Until recently, typical phases reported included, e.g., face-centered cubic (FCC) for spheres[15], nematic and smectic liquid crystals for discs[16] and rods[17], plastic solids for ellipsoids[18], and parquet phases for parallelepipeds[19]. The computer simulation predictions of the self-assembly of a dodecagonal quasicrystal with regular tetrahedra[20] (and later triangular bipyramids[21]) suggested for the first time the complexity that could be achieved for thermodynamically self-assembled structures solely with hard interactions. Since then, simulation predictions of ordered phases of hard polyhedra self-assembled from the fluid phase include, e.g., cubic crystals from cubes[22], sheared body-centered cubic (BCC) from octahedra[12,22,23], and several complex crystals – including diamond, β-tin, and high pressure lithium – for a family of truncated tetrahedra[23].

The thermodynamic phase behavior of hard particles is determined solely by particle shape and can be understood through entropy maximization[24]. Even for systems of weakly interacting particles, shape controls assembly when excluded volume dominates energetic interactions. Flat facets tend to align hard particles to maximize entropy, resulting in effective directional entropic forces[23] or "bonds". As a result, self-assembled phases are often different than maximally dense packings[25-27]. Because this effective statistical force increases with surface contact or alignment area, suitably facetted polyhedra can organize into various mesophases[28] and complex crystals with low coordination numbers like diamond[23].

The concept of directional entropic forces and their relation to particle faceting suggests that particle shape could, perhaps in combination with other information, be used to predict assembled phases. To establish clear quantitative trends, however, requires data on many different shapes. Here we present simulations of the self-assembly of 145 different polyhedra, including all the Platonic, Archimedean, Catalan and Johnson solids, and a sampling of prisms, antiprisms, zonohedra, and other polyhedra. Thousands of particles of each shape were simulated in multiple runs using Monte Carlo computer simulations, and the phases that formed were analyzed and identified. We show that the majority of shapes assemble into ordered structures, the most complex of which is γ-brass with 52 particles in the unit cell. We demonstrate that the isoperimetric quotient – a measure of particle shape – and the number of neighbors in the first neighbor shell of the dense *fluid* can be combined to predict the structural class of the ordered phase.

**Self-assembling polyhedra**

Fig. 1 shows the 145 particles simulated, classified according to the structure(s) they assemble into repeatedly from the dense fluid. We group polyhedra into four general categories: (i) crystals, (ii) plastic crystals, (iii) liquid crystals, and (iv) disordered (glassy). The categories are further subdivided into classes based on the type of order and crystallographic symmetry. Ordered structures all form at densities between 49% and 63%, depending on particle shape. We first note the remarkable finding that a large fraction, 101/145 ≈ 70%, of the polyhedra simulated form at least one ordered phase on the time scale of our simulations. This number is surprisingly high given the wide range of shapes simulated, and demonstrates a strong propensity for order in systems of polyhedra, even in the absence of explicit attractive interactions.

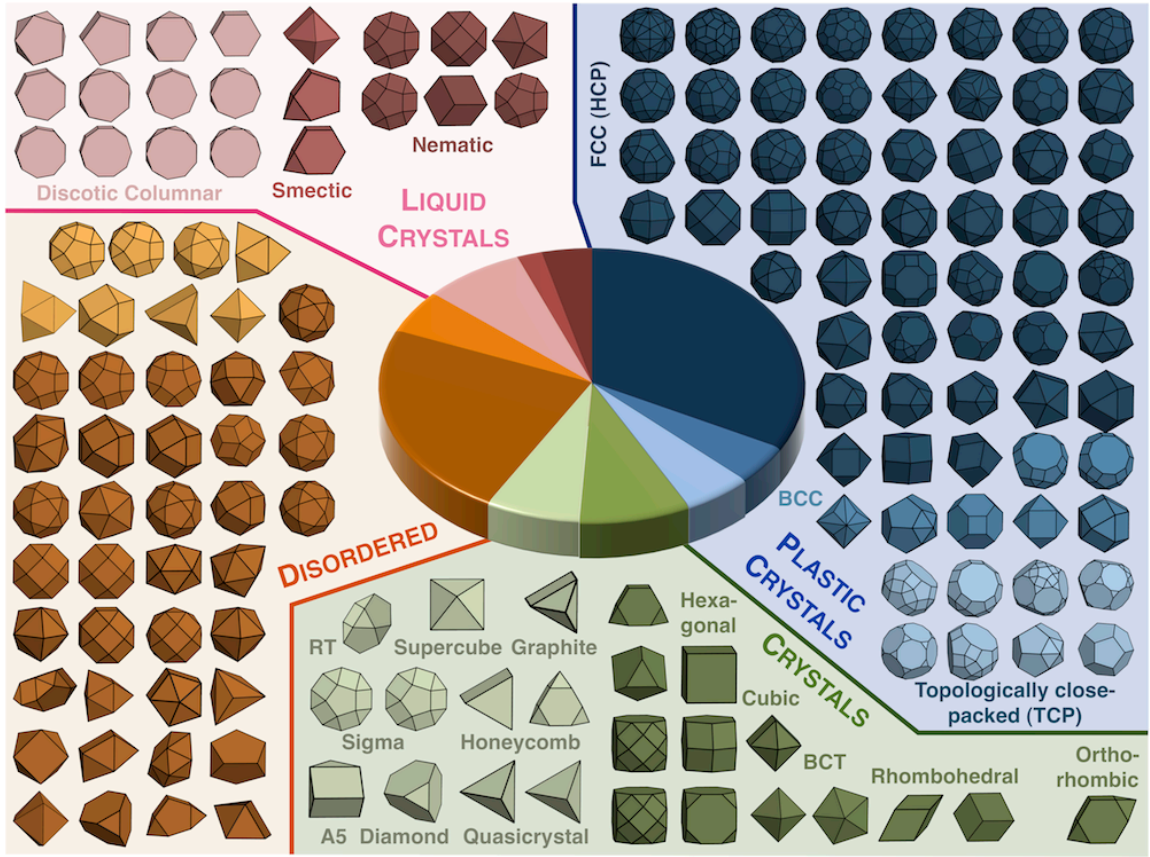

**Figure 1. Polyhedra and their assemblies.** In computer simulations, a majority of symmetric hard polyhedra is found to self-assemble from a disordered fluid phase into ordered phases. We distinguish four states of organization. Counterclockwise from lower right: crystals, plastic crystals, liquid crystals, and disordered (glassy). For the latter, no assembly is observed, and we distinguish those that strongly order locally with preferential face-to-face alignment (light orange) from those with only weak local order (dark orange). In the case of crystals, we distinguish Bravais lattices (dark green) and non-Bravais lattices (light green). The pie chart in the center compares the relative frequency of the observed classes of structures. In each class, polyhedra are listed in decreasing order of the isoperimetric quotient. A polyhedron is included multiple times if it was found to assemble into more than one ordered phase. The names of each polyhedron simulated can be found in Supplementary Information.

Fig. 2 demonstrates both the diversity and structural complexity of phases we find by showing representative phases assembled by 12 shapes. An in-depth discussion of the phase behavior for each of the 145 polyhedra is outside the scope of this work. Some additional information including the names of each polyhedron is found in Supplementary Information.

**Crystals.** For crystals, we find five different Bravais lattices (hexagonal, cubic, body-centered tetragonal (BCT), rhombohedral, orthorhombic) and several non-Bravais lattices. The lattice shear we observe with truncated cubes has been described in experiment[10,12,14]. The A5 lattice, graphite, honeycomb lattice, diamond structure[23], and supercube lattice are periodic and have only a few particles in the unit cell. The quasicrystals have been reported previously with tetrahedra[20] and triangular bipyramids[21]. A new type of hexagonal random tiling (RT) forming independent layers is observed for the bilunabirotunda (the Johnson solid J91). A two-dimensional version of the RT has recently been reported in a molecular network[29]. Unlike all other crystal structures we study which form large, sample-spanning crystals, we observe only small patches of the RT. It is conceivable that rhombic rectangular prisms, whose shape is close to the bilunabirotunda, form the same tiling faster and with fewer defects.

Four examples of crystals are analyzed in more detail in Fig. 2. Dürer's solids form a simple cubic crystal. The cubic crystal in Fig. 2a is special because it is a degenerate crystal[21,30]. Particle orientations align randomly along four equivalent orientations. The space-filling gyrobifastigium assembles into a crystal isostructural to β-Sn, the metallic form of tin (Fig. 2b). Pentagonal orthobicupola have a disk-like shape and arrange with their five-fold symmetry axes aligned into a crystal structure with the same tiling (3.4.3$^2$.4) as β-U. A periodic approximant to a dodecagonal quasicrystal, this tiling is known as the sigma-phase and has been observed in micelles[31,32] and colloids[33], but with different decoration of the tiles (Fig. 2h). Six square pyramids assemble into cubes ("supercubes") and then into a slightly sheared simple cubic lattice (Fig. 2i). The supercubes demonstrate the possibility of hierarchical assembly with hard particles, similarly to an FCC crystal reported for paired hemispheres[34].

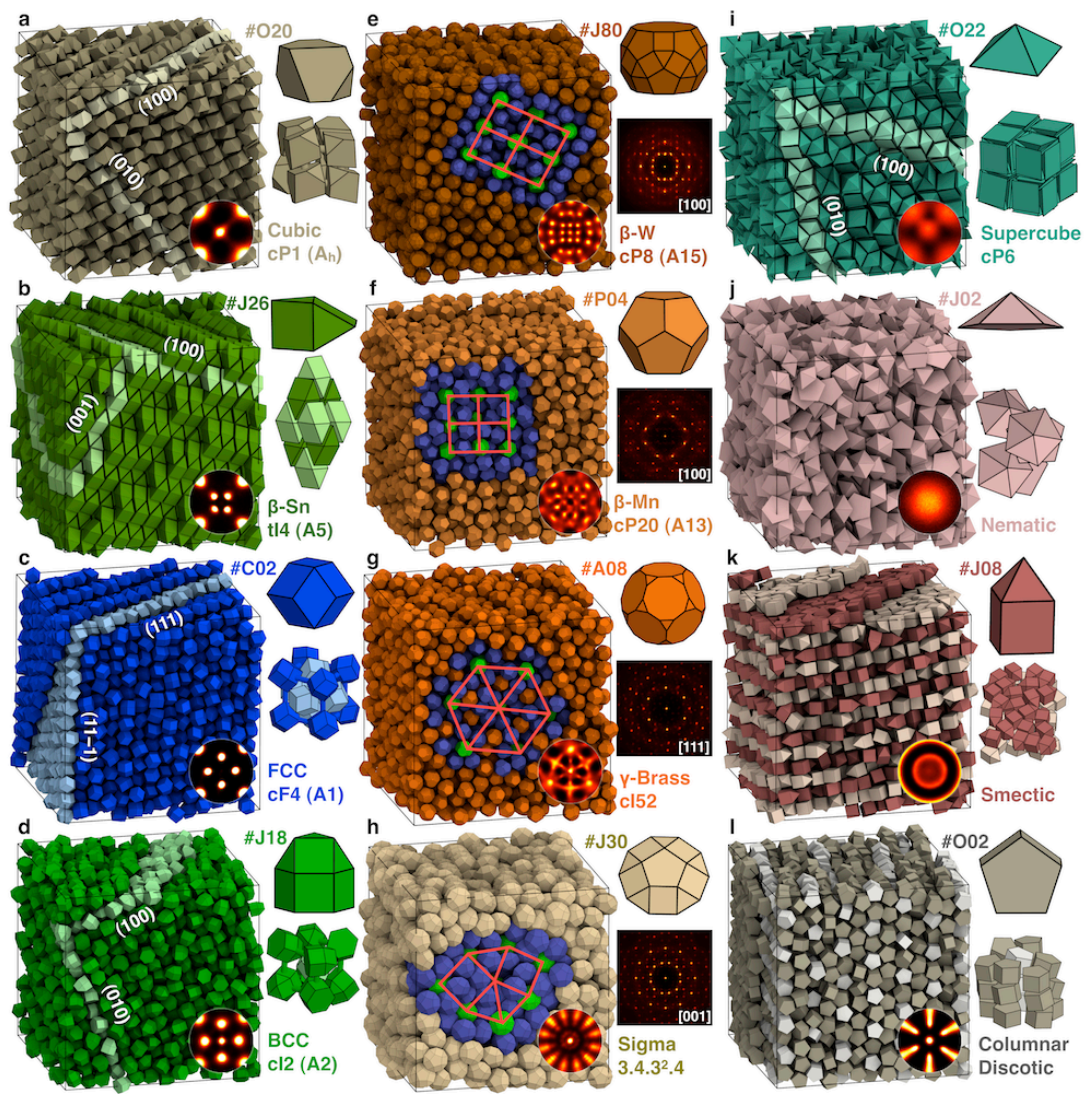

**Figure 2. Ordered phases of hard polyhedra.** Systems of 2048 polyhedra were assembled starting from the disordered fluid. In each subfigure, a snapshot of the simulation box (left), the bond-order diagram for nearest neighbors (inset), the polyhedron shape and ID (top right), a small group of particles or the diffraction pattern (middle right), and the crystallographic characterization consisting of name or atomic prototype, Pearson symbol and Strukturbericht designation (bottom right) are shown. The snapshots depict crystals (**a-b** and **h,i**), plastic crystals (**c-g**), and liquid crystals (**j-l**). Some low index planes (**a-d**), tilings descriptions consisting of squares and triangles (**e-h,i**) and structural features (**k,l**) are highlighted in the simulation snapshots by different colors.

**Plastic crystals.** We find that 66 of the 145 particle shapes crystallize into rotator phases known as plastic crystals. The plastic crystals we find all correspond to crystallographically dense packings, including FCC (or hexagonally close-packed, HCP) and BCC structures and three topologically close-packed (TCP) polytetrahedral structures isostructural to β-W, β-Mn, and γ-brass. We do not distinguish between FCC and HCP, because simulation results often have high densities of stacking faults. In a TCP structure lattice sites are coordinated by distorted tetrahedra – a generalization of the Frank-Kasper phases[35,36]. We always and exclusively observe plastic crystals for these three types of crystals and, in some cases, a diffusionless transformation to a non-rotator crystal at higher density. Although we did not yet study the phase behavior at high density, we expect all plastic crystals to eventually transform into a non-rotator phase.

In Fig. 2c we show that rhombic dodecahedra (the Voronoi cell of FCC) order into an FCC plastic crystal. We observe that the plastic phase transforms into a non-rotator phase at higher densities. Elongated triangular cupolas assemble a plastic BCC crystal (Fig. 2d). The formation of a high-symmetry phase is counter-intuitive given the asymmetric axial shape of the cupola. The paradiminished rhombicosidodecahedron has two large parallel faces and forms a plastic TCP phase isostructural to β-W (Fig. 2e). This phase, also known as the A15 structure, is frequently observed with micelles[37,38]. Dodecahedra assemble into the complex β-Mn structure (Fig. 2f). Since the distribution of Bragg peaks in the diffraction pattern resembles eight-fold symmetry, β-Mn can be interpreted as an approximant of an octagonal quasicrystal[39]. Indeed, we often observe eight-fold symmetry in the diffraction pattern during crystallization at an intermediate stage. With truncated dodecahedra we observe γ-brass (Fig. 2g). With 52 atoms per unit cell, this is the most complex periodic crystal observed in this study.

**Liquid crystals.** In the case of liquid crystals, we find nematic, smectic, and discotic phases. The pentagonal pyramid has a platelet-like shape and stabilizes a nematic liquid crystal (Fig. 2i). The up/down orientation of the pyramid relative to the director is random. Obtuse golden rhombohedra orient all of their axes completely and form a biaxial nematic (Fig. 2j). We observe a transition to a nematic liquid crystal at lower density. The elongated square pyramid arranges randomly in smectic layers (Fig. 2k). We confirmed there is no preferred orientation within the layers. Like all regular prisms and antiprisms with five-fold or higher symmetry, the pentagonal prism assembles a columnar phase (Fig. 2l)[40,41]. Particles are free to both shift along and rotate around the column axis.

**Disordered.** Some polyhedra are never observed to self-assemble into an ordered phase on the time scale of the simulation, despite run times more than an order of magnitude longer than that needed for the slowest crystal former in our study. Instead, we observe for 44 of the shapes in Fig. 1 that the particle dynamics becomes increasingly slow with increasing packing fraction, eventually producing a glassy or jammed state without rotational or translational order. Studies of dense packings of these shapes[27] yield crystals with packing fractions higher than that of the disordered phases achieved in our simulations. This suggests that these 44 particles are not intrinsic glass formers and instead have ordered "ground states" in the limit of infinite pressure. Similar to the tetrahedron[20] and triangular bipyramid[21], however, the infinite-pressure packing may be thermodynamically stable only at very high packing fractions and pressures where the kinetics are too slow to allow equilibration, even in experiments. Likewise, most atomic and molecular glass formers also have crystalline ground states that are difficult to access thermodynamically.

It is interesting to note that 41 of the 44 glass formers are Johnson solids and most, but not all, are non-centrally symmetric. Johnson solids are typically less symmetric than Platonic and Archimedean solids, which are all found to order in our simulations. This agrees with the intuition that highly symmetric polyhedra might be more easily assembled than non-symmetric ones. Moreover, several of those shapes resemble prolate ellipsoids, which are known to be hard to crystallize[42,43]. These polyhedra thus represent a new class of glass-forming systems that may provide additional insight into the role of shape in jamming and the glass transition.

**Local order and entropic bonding**

We find we can relate particle shape to local order for plastic crystals, crystals, and liquid crystals. In Fig. 3, we study six example shapes, two for each category. We use a shape descriptor[44] that is the distribution of the face normal vectors of the particle on the sphere, weighted by the face area. This is a good descriptor to characterize directional entropic forces because face normal vectors indicate the directions in which we expect the alignment of neighboring faces by entropic "bonds". A second shape descriptor – the bond order diagram of nearest neighbors – measures the actual direction of "bonds" as found during the simulation. Similarity of the two measures indicates a direct relation between particle shape and local order. The relative orientation of the neighbors is analyzed using orientational correlation diagrams (see Methods), which are sensitive to rotations that change the degree of alignment, but not to rotations around the bonds. Peaks are expected in the orientational correlation diagram if face-to-face alignment persists over time, and a uniform intensity if it does not.

Consider the plastic crystal in Fig. 3a. We find that the polyhedra that form plastic phases often have many small faces; consequently, the shape descriptor exhibits many peaks in all directions. Indeed, bond directions are only weakly correlated with face normals. Also, the orientational correlation diagram is isotropic, as expected if face-to-face alignment is not a factor. This shows that directional entropic forces are not important for plastic crystals. To first approximation, polyhedra that form plastic crystals assemble into crystal structures known to be good sphere packings. Highly spherical particles form the densest sphere packings FCC and HCP. TCP phases are a compromise between high density and maintaining icosahedral local order present in the dense liquid[40]. The coordination geometry can be visualized with Voronoi cells (Fig. 3d). Voronoi cells of TCP phases often have pentagonal or hexagonal faces. Indeed, we frequently find TCP phases with particles that resemble the Voronoi cells, such as the (truncated) dodecahedron.

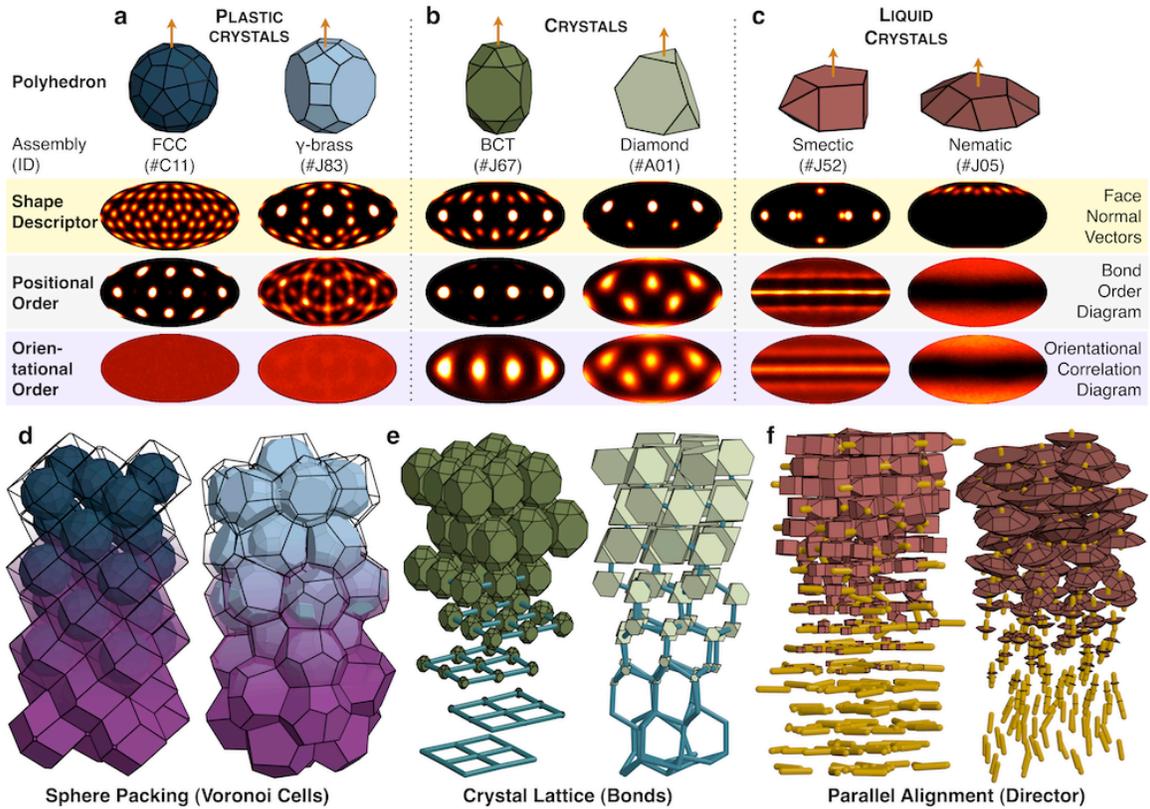

**Figure 3. Characterization of crystalline order.** We analyze six systems of polyhedra that span the three states of order observed with hard polyhedra. **a-c**, Polyhedra and their IDs are shown at top. The distribution of face normal vectors, the bond order diagram, and orientational correlation diagrams (the latter two for nearest neighbors) are measures for the shape, positional local order and orientational local order, respectively. The diagrams demonstrate that entropic bonds are weak in plastic crystals (**a**), strong in crystals (**b**), and highly axial in liquid crystals (**c**). **d-f**, Small groups of particles are extracted from simulation snapshots. In plastic crystals, polyhedra rotate inside the Voronoi cells (**d**). Bonds in the direction of the face normal are important for crystals (**e**). Parallel arrangement dominates in the case of liquid crystals (**f**). From top to bottom, the transparency of Voronoi cells is decreased and/or the size of polyhedra is reduced.

Polyhedra that form crystals (Fig. 3b) are more aspherical with more pronounced, and fewer, faces. We find a strong correlation between the bond order diagram and the direction of face normal vectors, which demonstrates that face-to-face alignments dominate. Orientational correlation diagrams show strong peaks confirming pronounced directionality. The crystal lattice is thus well represented by an ordered network of entropic "bonds" (Fig. 3e). We observe that polyhedra assembling into crystals do not always resemble the Voronoi cells of the crystal (Fig. 2a).

Polyhedra forming liquid crystals typically have an axial shape, which is reflected in the shape descriptor by the dominance of a few faces (Fig. 3c). Bond order diagrams and orientational correlation diagrams of nematics and smectics exhibit continuous rotational symmetry. We observe that particles can form bonds in certain directions, but often rotate freely around these bonds. Alignment of the most prominent faces is important and can be analyzed by the alignment of the directors (Fig. 3f). We note that even though most of the polyhedra we study that form liquid crystals are oblate (Fig. 1), the same considerations should hold for highly prolate shapes. In general, we expect for axial particles to align prominent faces and long particle dimensions first. We can also distinguish trends among liquid crystals. Shapes forming discotic liquid crystals are dominated by two parallel faces, which induces columnar stacking. In smectic phases, the rotation around the prominent faces is frustrated and, rather than columns, the particles instead order into layers, where they can freely translate. In nematic liquid crystals, stacking is further disfavored by the lack of prominent parallel faces and the particles merely align without translational order, exchanging rotational degrees of freedom for translational entropy.

**Towards structure prediction**

The vast amounts of data obtained on the large number of different polyhedra we study here allow us to test whether predictive rules exist relating shape and structure. The shape of a polyhedron can be described by the radially averaged shape density profile, which we relate to the structural order of the assembly in Fig. 4. The shape profile is a step function for a sphere and decays rapidly for highly spherical shapes (Fig. 4a), which typically form plastic crystals. Polyhedra with pronounced faces have a slower (intermediate) decay and form crystals (Fig. 4b). Since polyhedra assembling into liquid crystals have the most facetted and most asymmetric shapes, their shape profile decays slowest (Fig. 4c).

Next, we extract two scalar observables to further relate shape to structure. Several parameters have been used in the literature to analyze the shape of polyhedra. These include shape sphericity[25,27], shape anisotropy[28,45], and rotational symmetry[28]. A parameter that is sufficiently sensitive to large shape changes, but not too sensitive to small deformations, is the isoperimetric quotient, defined as $IQ = 36\pi V^2/S^3$, where $V$ is volume and $S$ is surface area[45–47]. $IQ$ has the advantage that it is does not depend on the definition of a center (e.g. the centroid) and is scale-invariant. A second scalar parameter accounts for the configuration of bonds in the ordered phase. We choose the number of nearest neighbors, $N_o$, in the ordered phase. In the case of glasses, we measure the number of nearest neighbors at density 55%, which is the density where we typically observe crystallization of polyhedra that do not form glasses.

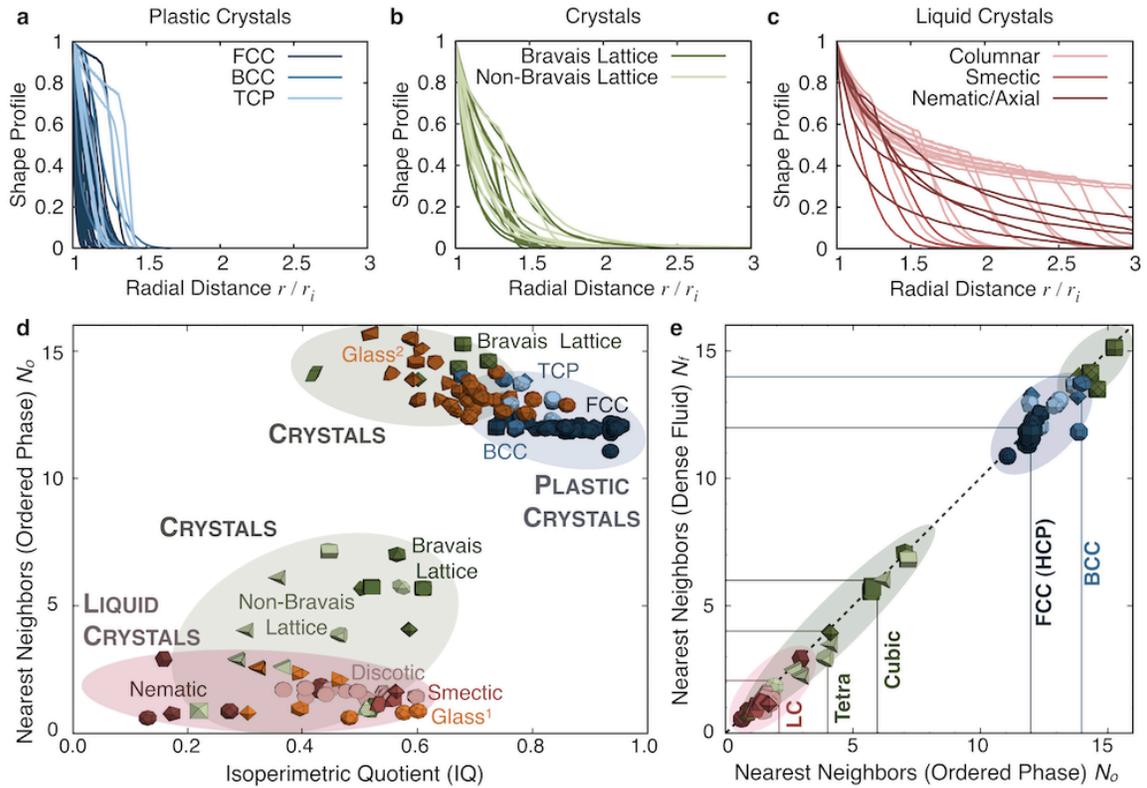

**Figure 4. Correlation between shape and assembly behavior. a-c**, The radial shape profile decays most rapidly for plastic crystals, less rapidly for crystals, and least rapidly for liquid crystals, indicating that polyhedra forming plastic and liquid crystals are typically spherical and highly asymmetric, respectively. **d**, The number of nearest neighbors in the ordered (or glass, see text) phase, $N_o$, is correlated to the isoperimetric quotient (*IQ*) of the polyhedron. Here, *IQ* is a scalar parameter for the sphericity of the shape and the number of neighbors is a measure for the degree of local order. Data points are drawn as small polyhedra. Polyhedra are colored and grouped according to the assemblies they form. **e**, Polyhedra have, in most cases, nearly identical numbers of nearest neighbors in the ordered phase ($N_o$) and the fluid phase ($N_f$) close to the ordering transition. Because of this strong correlation, combining $N_f$ and *IQ* allows for prediction of the class of ordered structure expected for most cases.

A clear correlation between *IQ* and $N_o$ can be seen in Fig. 4d. We relate the parameters to the class of order found in simulation. From *IQ* alone, we can predict that highly spherical particles (0.8 ≤ *IQ* < 1.0) form plastic crystals, and highly aspherical particles (*IQ* < 0.3) are candidates for nematic liquid crystals. By combining *IQ* and $N_o$, we observe that polyhedra forming crystals appear in two groups separated by a gap, one group at intermediate *IQ* and high $N_o$, another group at intermediate *IQ* and intermediate $N_o$. Phases with few nearest neighbors ($N_o$ < 3) often form liquid crystals. Thus although plastic crystals and nematic liquid crystals can be anticipated solely by the *IQ* value of the polyhedron, predicting other ordered phases requires additional information contained in $N_o$, with the exception of hierarchical crystals like that in Fig. 2i, for which $N_o$ should be calculated differently.

Remarkably, this additional information can also be obtained from the disordered fluid phase. We compare the number of nearest neighbors measured close to the onset of ordering in the fluid ($N_f$) and in the ordered phase ($N_o$) in Fig. 4e. We find that both numbers are nearly identical for almost all 145 shapes. Consequently, by combining the isoperimetric quotient and the average number of neighbors in the dense fluid – a number easy to obtain in short simulations and experiments by integrating over the first peak of the radial distribution function – we can predict with reasonable accuracy the class of structure that will form from the fluid.

When comparing our observations with known crystal structures of atoms and molecules, which can be rationalized in terms of a few parameters like the strength and directionality of bonds between atoms[48] and the molecular geometry[49], we can interpret our findings as follows. First, FCC (HCP), BCC, and TCP crystals formed from spherical hard polyhedra have non-directional or weakly directional entropic interactions. Their assembly is dominated by packing, and their atomic analogue is metals and metallic bonding. It is interesting to note that all of our plastic crystals except gamma-brass are found in elementary metals. Second, complex crystals have strong directional entropic bonding, reminiscent of covalent bonds. Third, liquid crystals are dominated by the asymmetry of the particle shape. They correspond in their behavior most closely to molecular liquid crystals.

Our results push the envelope of entropic crystallization and the assembly behavior of hard particle fluids. Quantitative analysis of the effective strength and directionality of entropic forces is necessary to set our findings on a firm theoretical base. Analogous to recent progress in the structure prediction of crystals made from atoms[50], a more precise, robust framework able to predict the structure into which a given polyhedron will self-assemble is desired. Beyond hard particles, our results may provide insight into the role of shape in the crystallization of weakly interacting systems of colloids, nanoparticles, proteins and viruses.

## Acknowledgements

This material is based upon work supported by the DOD (N00244-09-1-0062). ME acknowledges support from the Deutsche Forschungsgemeinschaft (EN 905-1/1). We thank R.G. Petschek and J.M. Millunchick for comments on the manuscript.

## Methods

**Polyhedra.** The 145 convex polyhedra included in this study were chosen for their high symmetry. They are the Platonic solids (#P01-P05, regular and identical faces), Archimedean solids (#A01-A13, regular faces with identical vertices), Catalan solids (#C01-C13, dual of the Archimedean solids), Johnson solids (#J01-J92, regular faces without identical faces), and other symmetric polyhedra (#O01-O22, including prisms, antiprisms and zonohedra). In the case of prisms and antiprisms the edge length is chosen to be the same for all edges. A detailed list of the polyhedra and some of their geometric and assembly properties is given in Supplementary Information.

**Computer simulations.** Monte Carlo simulations were conducted to equilibrate systems of hard polyhedra using isochoric simulations similar to Ref.[23]. The polyhedra overlap was determined exactly using the Gilbert-Johnson-Keerthi algorithm. For each polyhedron, an initial set of 12 simulations with 2048 polyhedra was run at packing densities 49% ≤ $\phi$ ≤ 60% for 3·10$^7$ MC cycles. Most polyhedra assembled under these conditions. If self-assembly was not successful, then we ran longer simulations of 10$^8$ MC cycles and densities up to 63%. If still no crystallization was observed, the system was labeled "disordered". For the complex topologically close-packed phases, larger systems of up to 10000 particles were studied to determine the crystal structure unambiguously.

**Analysis of local and global order.** We define *nearest neighbors* as pairs of polyhedra corresponding to the first peak of the radial distribution function calculated from the centroids. In the same sense, *Voronoi cells* are calculated using centroids. Fourier transforming the centroids generates *diffraction patterns*. *Bonds* connect nearest neighbors. Given polyhedra with centroids $r_i$ and rotation matrix $M_i$, the *bond order diagram* is the distribution of bonds $\{r_j - r_i, i \text{ bonded to } j\}$ projected on the surface of a sphere. Analogue to the bond order diagram, we define the *orientational correlation diagram* as the distributions of the vectors $\{M_j M_i^{-1}(r_j - r_i), i \text{ bonded to } j\}$ on the surface of a sphere. The orientational correlation diagram is sensitive to the rotation of pairs of neighbors around the bond connecting their centroids.

**Crystal structure determination.** To identify a crystal structure, atoms are positioned at the centroids of polyhedra. We consider this replacement when we say that an arrangement of polyhedra is isostructural to an atomic crystal. In a first step, crystal structures were analyzed in real space using particle positions and bond-order diagrams. This allows determining whether a system is ordered or disor-dered. We distinguish plastic (rotator) crystals from non-rotator crystals by the absence of long-range correlations in the particle-bond and particle-particle correlation diagram. Liquid crystals are identified by the absence of Bragg peaks in the diffraction pattern along at least one spatial direction. In the case of topologically close-packed phases, the crystal structure is too complicated to analyze in real space and we compare diffraction patterns of simulation results with those of ideal crystal lattices.

**Characterization of polyhedra shape.** In the diagrams showing the *distribution of face normal vectors*, Gaussians with fixed width and intensity proportional to the area of a polyhedron face are positioned in the direction of all face normals. The *isoperimetric quotient Q* of a polyhedron $P$ with volume $V$ and surface area $S$ is the scalar defined as $Q = 36\pi V^2/S^3$. The *radial shape profile s(r)* is the fraction of a sphere surface that is contained inside the polyhedron, $s(r) = (4\pi)^{-1} \oiint \chi_P(r,\theta,\phi) \sin\theta \, d\theta d\phi$, where $\chi_P$ is the characteristic function of the polyhedron with origin at the centroid. The shape profile lies between 0 and 1, and has the property $V = \int_0^\infty 4\pi r^2 \, s(r) \, dr$.

**Visualization.** Simulation snapshots are rendered using Phong shading and ambient occlusion to improve depth perception. Functions defined on the surface of a sphere (bond order diagram, correlation diagrams) are shown in either stereographic projection (angle preserving, circular plots) or using the Mollweide projection (area preserving, elliptical plots). These functions and the diffraction patterns are temporally averaged over short simulation times to remove speckle patterns.